\documentclass[usenatbib,referee]{mn2e}

\usepackage{amsmath, amssymb}
\usepackage[dvips]{graphicx}
\usepackage{epsfig}

\newcommand{\Blight}{B_{\rm L}}
\newcommand{\rlight}{r_{\rm L}}

\newcommand{\rot}{\mathbf{\nabla} \times}
\newcommand{\divg}{\mathbf{\nabla}\cdot}

\newcommand{\ex}{\mathbf{e}_{\rm x}}
\newcommand{\ey}{\mathbf{e}_{\rm y}}
\newcommand{\ez}{\mathbf{e}_{\rm z}}

\newcommand{\aap}{A\&A}
\newcommand{\mnras}{MNRAS}

\newcommand{\apj}{ApJ}
\newcommand{\apjl}{ApJL}
\newcommand{\apjs}{ApJS}

\newcommand{\nat}{Nature}

\title[Polarization of the striped wind emission]{Phase-resolved polarization properties of the pulsar striped wind synchrotron emission.}

\author[J. P\'etri]{J. P\'etri$^{1}$\thanks{E-mail: jerome.petri@astro.unistra.fr}\\
  $^{1}$Observatoire Astronomique de Strasbourg, Universit\'e de Strasbourg, CNRS, UMR 7550, 11 rue de l'Universit\'e, 67000 Strasbourg, France.}

\begin{document}

\date{Accepted . Received ; in original form }

\maketitle

\begin{abstract} 
Since the launch of the Fermi telescope more than five years ago, many new gamma-ray pulsars have been discovered with intriguing properties challenging our current understanding of pulsar physics. Observation of the Crab pulsar furnish today a broad band analysis of the pulsed spectrum with phase-resolved variability allowing to refine existing model to explain pulse shape, spectra and polarization properties. The latter gives inside into the geometry of the emitting region as well as on the structure of the magnetic field. Based on an exact analytical solution of the striped wind with finite current sheet thickness, we analyze in detail the phase-resolved polarization variability emanating from the synchrotron radiation. We assume that the main contribution to the wind emissivity comes from a thin transition layer where the dominant toroidal magnetic field reverses its polarity, the so-called current sheet. The resulting radiation is mostly linearly polarized. In the off-pulse region, the electric vector lies in the direction of the projection onto the plane of the sky of the rotation axis of the pulsar. This property is unique to the wind model and in good agreement with the Crab data. Other properties such as a reduced degree of polarization and a characteristic sweep of the polarization angle within the pulses are also reproduced. These properties are qualitatively unaffected by variations of the wind Lorentz factor, the lepton injection power law index, the contrast in hot and cold particle, the obliquity of the pulsar and the inclination of the line of sight.
\end{abstract}


\section{INTRODUCTION}

With the advent of the Fermi/LAT, more than one hundred new gamma-ray pulsars have been detected \citep{2012ApJS..199...31N}. Their spectral properties rule out a polar cap scenario for pulsed high-energy emission because the cut-off frequency about a few GeV is only exponential and not super-exponential as predicted by these models~\citep{2010ApJS..187..460A}. This favours emission sites in the outer part of the magnetosphere or even outside the light-cylinder. However, with our current knowledge relying mostly on phase-resolved spectra, it is impossible to go further and eliminate some of the remaining emission geometries such as outer gaps, two-pole caustic slot gaps \citep{2004ApJ...606.1125D} and the striped wind \citep{2005ApJ...627L..37P}. This requires additional informations about, for instance, the magnetic field configuration at the location where the high-energy photons are produced. The polarization properties of the pulsed emission could give us a hint about the magnetic field and discriminate between competing scenarios. Although there are only very little data available in optical or shorter wavelengths, polarization measurements will put severe constraints on these models. 

Polarization properties at multi-wavelength would certainly help to discriminate between several geometries of the sites of emission. Gamma-ray light-curves alone can already give good insight into the magnetosphere as shown by \cite{2010ApJ...714..810R} or about the geometry of the striped wind \citep{2011MNRAS.412.1870P}. The new sample of Fermi-LAT gamma-ray pulsars increased interest into modelling of gamma-ray emission. Detection of several millisecond gamma-ray pulsars was not expected and came as a real surprise.  Thus, \cite{2009ApJ...707..800V} focused special attention to this class of millisecond pulsars to probe the geometry of the emission regions, taking into account relativistic effects.

Because of its brightness, the Crab pulsar is the most extensively studied pulsar in the literature and therefore the best candidate to diagnose the magnetic field geometry through the detection of polarized emission. Because it is one of the youngest gamma-ray pulsars and showing high brightness and luminosity, many precise measurements have been done, ranging from radio observations through optical up to X-rays and gamma-rays. Highly accurate fully phase-resolved optical observations performed with the OPTIMA instruments have been done in the past by \cite{2009MNRAS.397..103S}, completing earlier works by \cite{1969ApJ...157L...1W, 1970Natur.227.1327C, 1970ApJ...162..475K, 1981MNRAS.196..943J} and \cite{1988MNRAS.233..305S}. These studies motivated comparative studies of the emission from polar caps, outer gaps \cite{1995ApJ...438..314R} and two-pole caustic models \citep{2003ApJ...598.1201D,2004ApJ...606.1125D}.  In all of these models, the radiation is produced within the light cylinder. However, the pulse profile is determined by the assumed geometry of the magnetic field and the location of the gaps. Neither of these models are able to fit the optical polarization properties of the Crab pulsar. They fail to explain the observed constant value of the polarization angle in the off-pulse state. The phase alignment between radio pulses and optical has also been checked recently. At shorter wavelengths, the data become more sparse. For instance, in UV the Crab was detected by \cite{1996MNRAS.282.1354G} in X-rays/soft gamma rays \citep{1978ApJ...225..221S}, hard X-rays \citep{2008ApJ...688L..29F} and in gamma rays \citep{2008Sci...321.1183D}. There exists only a handful of pulsars for which polarization has been reported, exclusively in optical and mostly a phase-averaged value can be extracted.

Models of optical polarization of the Crab pulsar have been proposed early after its discovery. \cite{1973ApJ...183..977F} introduced a relativistic vector model which has been fitted to the Crab by \cite{1973ApJ...183..987C} and improved by subsequently more accurate measurements by \cite{1974ApJ...190..375F}.

The spectral features above the optical are clearly non thermal and can be explained via some synchrotron and inverse Compton radiation. However, the detailed mechanism, the geometry and location of the site of emission for this high-energy, pulsed emission is poorly constrained. It is usually explained in the framework of either the polar cap~\citep{1970Natur.227..465S, 1975ApJ...196...51R} or the outer gap models~\citep{1986ApJ...300..500C}. Nevertheless, these models still suffer from the lack of a self-consistent solution for the pulsar magnetosphere and are based on the assumption that the magnetic field structure is that of a rotating dipole.

It seems reasonable to fit the pulsed emission from the Crab pulsar by a two-component model with a precise track of the pulse shape evolution with increasing energy \citep{2006A&A...459..859M}. The detection of pulses above 25~GeV by MAGIC \citep{2008Sci...322.1221A} motivated further investigations \citep{2009A&A...499..847C}. \cite{2007ApJ...670..677T} undertook a detailed analysis of the polarization and phase-resolved curvature, synchrotron and inverse Compton emission from the outer gap model. This study has recently been refined by \cite{2008ApJ...676..562T}.  Some other authors invoked an anisotropic synchrotron model for infra-red to X-ray emission in the framework of the outer gap scenario, see \cite{2001ApJ...546..401C}.

In this paper, we pursue our effort to investigate the ability of an alternative site for the production of pulsed radiation as firstly demonstrated by \cite{2002A&A...388L..29K}, based on the idea of a striped pulsar wind, originally introduced by \cite{1990ApJ...349..538C} and \cite{1994ApJ...431..397M} and further elaborated by \cite{2001ApJ...547..437L} and \cite{2003ApJ...591..366K}.  The striped wind model gave already satisfactory fits to the optical polarization data from, for instance, the Crab pulsar~\citep{2005ApJ...627L..37P} assuming synchrotron radiation. By extension to higher energies due to inverse Compton scattering, it is also possible to fit the pulsed and spectral variability of the gamma-ray spectra of the Geminga pulsar with reasonable accuracy \citep{2009A&A...503...13P}. Synchrotron emission is able to reproduce the sample of Fermi gamma-ray pulsars as shown in \cite{2012MNRAS.424.2023P}. For a throughout history of emission mechanisms outside the light-cylinder, the reader is referred to the introduction of the above mentioned papers to find more literature on this topic. Here, we compute light-curves as well as phase-resolved polarization properties of gamma-ray pulsars in the synchrotron regime. 
Radiation is mainly produced in the relativistically hot and dense current sheet by the synchrotron process. We use an exact analytical solution to the time dependent Maxwell equations for a finite thickness of the stripes, attempted to relax the infinitely thin current sheet assumption. Details of the model, including closed expressions for the electromagnetic field are given in Sec.~\ref{sec:Modele}. We then compute the properties of the synchrotron polarization characteristics. The relevant results are discussed in Sec.~\ref{sec:Results} before concluding this study.

\section{THE STRIPED WIND MODEL}
\label{sec:Modele}

In this section, we present an improved analytical model of the striped wind with finite current sheet thickness. Its main advantage compared to the prescription of \cite{2005ApJ...627L..37P} is that no extra polo\"idal component~$B_\vartheta$ has to be introduced and the electromagnetic field satisfies the homogeneous Maxwell equations exactly.

\subsection{Magnetic field structure}

It is indeed possible to get an exact analytical solution for the electromagnetic field in a realistic striped pulsar wind, i.e. with a finite thickness for the current sheet and a magnetized flow wind expanding radially outwards at a constant speed~$V=\beta_{\rm v}\,c$ slightly less than the speed of light denoted by~$c$. It is straightforward to check that the following magnetic field structure satisfies the homogeneous set of Maxwell's equation. In spherical polar coordinates $(r,\vartheta,\varphi)$ centred on the star and with the $z$-axis along the rotation axis, the explicit expressions for the electromagnetic field components in the rest frame of the star are
\begin{subequations}
  \label{eq:Champ_EM_Strie}
\begin{align}
  B_r & = \beta_{\rm v}^2 \, \Blight \, \frac{\rlight^2}{r^2} \, \tanh (\Psi_s/\Delta) \\
  B_\vartheta & = 0 \\
  B_\varphi & = - \beta_{\rm v} \, \Blight \, \frac{\rlight}{r} \, \sin \vartheta \, \tanh (\Psi_s/\Delta) \\
  E_r & = 0 \\
  E_\vartheta & = - \beta_{\rm v}^2 \, c \, \Blight \, \frac{\rlight}{r} \, \sin \vartheta \, \tanh (\Psi_s/\Delta) \\
  E_\varphi & = 0 
\end{align}
\end{subequations}
Here, $\rlight=c/\Omega$ is the radius of the light cylinder, $\Omega$ is the angular velocity of the pulsar, $\Blight$ is a fiducial magnetic field strength in the vicinity of the light cylinder, $V$ is the radial speed of the wind and $\Delta$ represents a parameter quantifying the length scale of the current sheet thickness. It is easily checked that the homogeneous Maxwell equations, namely $\rot \mathbf{E} = - \partial_t \mathbf{B}$ and $\divg \mathbf{B}=0$ are exactly and analytically satisfied by the solution in eq.~(\ref{eq:Champ_EM_Strie}). The current sheet is located in regions where the function
\begin{equation}
  \label{eq:PSI_S}
  \Psi_s = \cos \vartheta \, \cos \chi + \sin \vartheta \, \sin \chi \, \cos\left[\varphi - \Omega \, ( t - \frac{r}{\beta_{\rm v}\,c} ) \right]
\end{equation}
is nearly zero, $\chi$ is the obliquity, i.e. the angle between the magnetic and rotation axes. Note also that the magnetic structure does not possess the property $B_r = B_\varphi$ in the equatorial plane of the light-cylinder. The ratio of their magnitude at that point is equal to~$\beta_{\rm v}=V/c$. Nevertheless, this solution is physically satisfactory because it satisfies the constrain $E<c\,B$ everywhere in space. Indeed, the electric drift speed is given by
\begin{subequations}
  \label{eq:Vitesse_Derive_Electrique}
 \begin{align}
  v_r^E & = \beta_{\rm v} \, c \, \frac{\sin^2 \vartheta}{\sin^2 \vartheta + \beta_{\rm v}^2 \, \rlight^2/r^2} \\
  v_\vartheta^E & = 0 \\
  v_\varphi^E & = \beta_{\rm v}^2 \, c \, \frac{\rlight}{r} \, \frac{\sin \vartheta}{\sin^2 \vartheta + \beta_{\rm v}^2 \, \rlight^2/r^2}  
 \end{align}
\end{subequations}
Thus its magnitude becomes
\begin{equation}
  \label{eq:Derive_Electrique_Intensite}
  v_E = \beta_{\rm v} \, c \, \sqrt{ \frac{\sin^2 \vartheta}{\sin^2 \vartheta + \beta_{\rm v}^2 \, \rlight^2/r^2} }
\end{equation}
clearly less than $c$ for any $r>0$ and any $\vartheta$. As a consequence, there exists a reference frame for which the electric field vanishes, it is precisely the frame with speed equal to the electric drift speed~$v_E$. The usual synchrotron emissivity can thus be computed in this special frame with the wind density and magnetic field in the electric drift frame defined by the speed $v_E$, before switching to the observer frame via a Lorentz transform. The magnetic field strength in the electric drift frame is most easily deduced from the relativistic invariance of the quantity $E^2-c^2\,B^2$. Therefore we get
\begin{equation}
  \label{eq:Champ_Magnetique_Propre}
  {B'}^2 = B^2-\frac{E^2}{c^2} = \beta_{\rm v}^2 \, \Blight^2 \, \frac{\rlight^2}{r^2} \, \left[ \frac{\sin^2\vartheta}{\Gamma_{\rm v}^2} + \beta_{\rm v}^2 \, \frac{\rlight^2}{r^2} \right] \, \tanh^2 (\Psi_s/\Delta)
\end{equation}
primed quantities are measured in the electric drift frame of the wind. The model employed to compute the gamma-ray pulse shape emanating from the striped wind is briefly examined in this section. The geometrical configuration and emitting particle distribution functions follows the same lines as those described in \cite{2009A&A...503...13P}.  The magnetized neutron star rotates at an angular speed of~$\Omega$, thus a period of~$P=2\,\upi/\Omega$, directed along the $(Oz)$-axis, i.e. the rotation axis is given by $\mathbf{\Omega} = \Omega \, \ez$. We use a Cartesian coordinate system with coordinates~$(x,y,z)$ and orthonormal basis~$(\ex, \ey, \ez)$. The stellar magnetic moment is denoted by~$\boldsymbol\mu$, it is assumed to be a dipole and makes an angle~$\chi$ with respect to the rotation axis such that
\begin{equation}
  \label{eq:momamg}
  {\boldsymbol\mu} = \mu \, [ \sin\chi \, ( \cos (\Omega \, t) \, \ex + \sin (\Omega \, t) \, \ey ) + \cos\chi \, \ez ] .
\end{equation}
This angle is therefore defined by $\cos\chi = {\boldsymbol\mu} \cdot \ez / \mu$. The inclination of the line of sight with respect to the rotational axis, and defined by the unit vector $\mathbf{n}$, is denoted by the angles~$\{\zeta_1, \zeta_2\}$ in spherical polar coordinates.  Thus
\begin{equation}
  \mathbf{n} = \sin\zeta_1 \, \cos\zeta_2\,\ex + \sin\zeta_1 \, \sin\zeta_2 \, \ey + \cos\zeta_1 \, \ez .
\end{equation}
We have $\cos\zeta_1 = \mathbf{n} \cdot \ez$.

\subsection{Wind synchrotron emissivity}

Because the electric drift speed is less than the speed of light in whole space, it is possible to Lorentz transform the electromagnetic field from the observer to a frame where the electric field vanishes. In that new frame, we can use the usual synchrotron formula for the intensity and the Stokes parameters. In the wind frame where $\mathbf{E}'=0$, the magnetic field transforms as
\begin{equation}
  \mathbf{B}' = \dfrac{\mathbf{B}}{\Gamma_{\rm v}} + \dfrac{\Gamma_{\rm v}}{\Gamma_{\rm v}+1} \, ( {\boldsymbol\beta}_{\rm v} \cdot \mathbf{B} ) \, {\boldsymbol\beta}_{\rm v}
\end{equation}
Its intensity in this frame becomes
\begin{equation}
\label{eq:TLB}
 B'^2 = \frac{B^2}{\Gamma_{\rm v}^2} + ( {\boldsymbol\beta}_{\rm v} \cdot \mathbf{B} )^2
\end{equation}
which is exactly Eq.~(\ref{eq:Champ_Magnetique_Propre}). The aim of this paper is to show the behaviour of the pulsed high-energy light-curves emanating from the striped wind flow for different combinations of the magnetic field and emitting particles configurations. We will not perform a detailed study of the phase-resolved spectral variability. Thus for this purpose, it is sufficient to fix the particle density number and specify an index~$p$ for the power-law distribution in momentum space. We introduce a two component emission model including a relatively cold plasma part in the well organized magnetic field outside the stripe and a hot almost unmagnetized plasma part inside the stripe. We thus adopt the following expressions for the cold and hot particle density number respectively as
\begin{subequations}
  \label{eq:Densite}
\begin{align}
  n_{\rm cold}(\mathbf{r},t) & = \frac{N_{\rm c} \, {\rm tanh}^2 (\psi/\Delta)}{r^2} \\
  n_{\rm hot}(\mathbf{r},t) & = \frac{N_{\rm h} \, [ 1 - {\rm tanh}^2(\psi/\Delta)]}{r^2} 
\end{align}
\end{subequations}
$N_{\rm c}$ sets the particle density number outside the current sheet, whereas $N_{\rm h}$ defines the density inside the sheet. We refer to \cite{2009A&A...503...13P} for more details about the justification of this choice. The radial motion of the wind at constant speed and the conservation law of the particle number imposes an overall $1/r^2$ dependence on this quantity, which is further modulated because the energization occurs primarily in the current sheet mainly due to adiabatic cooling.

For the synchrotron emissivity, we assume an isotropic power law distribution of leptons, and use the delta function approximation in the electric drift frame thus
\begin{equation}
  \label{eq:Emissivite_Sync}
  j'_{\rm sync}(\varepsilon') = \frac{\sigma_T\,c}{6\,\upi} \, U_B' \, \frac{\gamma'^3}{\varepsilon'} \, n'_e(\gamma')
\end{equation}
where $n'_e(\gamma') = K_e(\mathbf{r},t) \, \gamma'^{-p}$, $\varepsilon'$ is the photon energy in this frame, $\sigma_T$ the Thomson cross section and $U_B'=B'^2/2\,\mu_0$ the magnetic energy density in this frame. We assume the emission commences when the wind crosses the surface $r=r_0\gtrsim \rlight$.

\subsection{Stokes parameters}

For synchrotron radiation, the electric vector is directed towards the direction given by~$\mathbf{n} \times \mathbf{B}$ where $\mathbf{n}$ is the unit vector along the line of sight and $\mathbf{B}$ the local magnetic field at the emission site. In order to relate the observations to the polarization properties in the wind, we have to find the direction of the polarization vectors in the lab frame~$K$ with respect to those in the drift frame~$K_{\rm v}$. Relativistic kinematic implies a rotation effect on the polarization trihedron when transforming from~$K_{\rm v}$ to $K$. For later convenience, we denote vectors of the electromagnetic wave with lower cases $(\mathbf{e},\mathbf{b})$ whereas fields imposed by the exterior, which is independent of the existence or not of this wave, will be marked by upper cases~$(\mathbf{E},\mathbf{B})$.

Note first that in the comoving frame~$K_{\rm v}$, the electric field of the linearly polarized wave is directed along a direction perpendicular to the magnetic field~$\mathbf{e}_{\rm B'}$ and to the line of sight~$\mathbf{n}'$, direction symbolized by the unit vector~$\mathbf{e}'_{\rm p} = \mathbf{n}' \wedge \mathbf{e}_{\rm B'}$ such that $\mathbf{e}' \propto \mathbf{e}'_{\rm p}$. Indeed, for an isotropic distribution of particles, the elliptic polarization vanishes by average over polarization in opposite directions. Only a linear polarization perpendicular to the magnetic field remains. The magnetic field of this wave is therefore $\mathbf{b}'=\mathbf{n}'\wedge \mathbf{e}'/c$. For a distant observer, at rest in~$K$, the measured electric field is given by
\begin{equation}
  \mathbf{e} = \Gamma_{\rm v} \, \left[ ( 1 + {\boldsymbol\beta}_{\rm v} \cdot \mathbf{n}' ) \, \mathbf{e}' - ( {\boldsymbol\beta}_{\rm v} \cdot \mathbf{e}' ) \, \left( \mathbf{n}' + \dfrac{\Gamma_{\rm v}}{\Gamma_{\rm v}+1} \, {\boldsymbol\beta}_{\rm v} \right) \right]
\end{equation}
This field has to be expressed in the observer frame. First we find
\begin{equation}
 \mathbf{e} = \Gamma_{\rm v} \, \mathcal{D}_{\rm v}^2 \, \mathbf{n} \wedge \left[ \mathbf{B}\,' + \mathbf{n} \wedge ( {\boldsymbol\beta}_{\rm v} \wedge \mathbf{B}\,' ) - \dfrac{\Gamma_{\rm v}}{\Gamma_{\rm v}+1} \, ( {\boldsymbol\beta}_{\rm v} \cdot \mathbf{B}\,' ) \, {\boldsymbol\beta}_{\rm v} \right]
\end{equation}
Then by replacing~$\mathbf{B}'$, introducing the vector $\mathbf{q}$ we arrive at
\begin{subequations}
  \label{eq:Pulse:VecQ}
 \begin{align}
 \mathbf{e} & = \mathbf{n} \wedge \mathbf{q}  \\
  \mathbf{q} & = c \, \mathcal{D}_{\rm v}^2 \, [ ( 1 - {\boldsymbol\beta}_{\rm v} \cdot \mathbf{n} ) \, \mathbf{B} + ( \mathbf{B} \cdot \mathbf{n} ) \, {\boldsymbol\beta}_{\rm v} ]  
 \end{align}
\end{subequations}
 $\mathcal{D}_{\rm v}$ is the Doppler boosting factor
\begin{equation}
 \mathcal{D}_{\rm v} = \frac{1}{\Gamma_{\rm v} \, ( 1 - {\boldsymbol\beta} \cdot \mathbf{n} )} .
\end{equation}
We used the formula of light aberration for the line of sight and eq.~(\ref{eq:TLB}) for the magnetic field. The norm of the electric field vector is~$\sqrt{q^2 - ( \mathbf{n} \cdot \mathbf{q} )^2}$. The unit linear polarization vector is then
\begin{eqnarray}
 \mathbf{e}_{\rm p} & = & \dfrac{\mathbf{n} \wedge \mathbf{q}}{\sqrt{q^2 - ( \mathbf{n} \cdot \mathbf{q} )^2}}
\end{eqnarray}
This result is valid in general, whatever the inclination between magnetic field and plasma speed, see \cite{2003ApJ...597..998L} for more details. The polarization trihedron denoted by $\{\mathbf{\boldsymbol\epsilon}_1, \mathbf{\boldsymbol\epsilon}_2, \mathbf{n}\}$ is chosen such that $\mathbf{\boldsymbol\epsilon_1}$ points in the direction parallel to the projection of the rotation axis~$\mathbf{e}_\Omega$ onto the plane of the sky therefore
\begin{eqnarray}
 \mathbf{\boldsymbol\epsilon}_1 & = & \frac{\mathbf{e}_\Omega - (\mathbf{e}_\Omega\cdot\mathbf{n}) \, \mathbf{n}}{||\mathbf{e}_\Omega - (\mathbf{e}_\Omega\cdot\mathbf{n}) \, \mathbf{n}||} \\
 \mathbf{\boldsymbol\epsilon}_2 & = & \mathbf{n} \times \mathbf{\boldsymbol\epsilon}_1
\end{eqnarray}
The polarization angle~$\psi$ measured by an observer in the rest frame~$K$ is chosen as the angle formed between the projection of the pulsar rotation axis onto the plane of the sky and the projection of the wave electric field onto the same plane. This angle~$\psi$ is defined by
\begin{subequations}
  \label{eq:Pulse:AnglePolDef}
 \begin{align}
  \cos\chi & = \mathbf{e}_{\rm p} \cdot \mathbf{\boldsymbol\epsilon}_1 = - \frac{\mathbf{q} \cdot \mathbf{\boldsymbol\epsilon}_2}{e} \\
  \sin\chi & = \mathbf{e}_{\rm p} \cdot \mathbf{\boldsymbol\epsilon}_2 = \frac{\mathbf{q} \cdot \mathbf{\boldsymbol\epsilon}_1}{e}  
 \end{align}
\end{subequations}
In our geometry, the rotation axis is aligned with the $\ez$ coordinate thus $\mathbf{e}_\Omega = \ez$ and
\begin{equation}
 \mathbf{\boldsymbol\epsilon}_1 = \frac{\ez - \cos\zeta_1 \mathbf n}{\sin \zeta_1}
\end{equation}
In the special case of $\zeta_1=0$, we choose $\mathbf{\boldsymbol\epsilon}_1 = \ex$ and $\mathbf{\boldsymbol\epsilon}_2 = \ey$.

The calculation of the Stokes parameters as measured in the observer frame involves simply integrating the emissivity over the wind volume. However, this requires special care, because the Lorentz boost from the rest frame of the emitting plasma involves not only beaming and Doppler shift, but also a change in the polarization angle due to aberration effects. The above treatment handled all these subtleties. After straightforward but lengthy manipulations involving Lorentz transformations, we find the Stokes parameters as measured by an observer at time~$t_{\rm obs}$ by the following integrals
\begin{equation}
  \label{eq:StokesParameters}
  \left\{
    \begin{array}{c}
      I \\
      Q \\
      U
    \end{array}
  \right\} (t_{\rm obs}) = \int_{r_0}^{+\infty} \int_0^{\upi} \int_0^{2\,\upi}
  s_0(\mathbf{r},t_{\rm ret}) \,  
  \left\{
    \begin{array}{c}
      \frac{p+7/3}{p+1} \\
      \cos \, (2\,\psi) \\
      \sin \, (2\,\psi)
    \end{array}
  \right\}
  \, r^2 \, \sin\vartheta \, dr \, d\vartheta \, d\varphi
\end{equation}
where the retarded time is given by $t_{\rm ret} = t_{\rm obs} + \mathbf{n}\cdot\mathbf{r}/c$. In this approximation the circular polarization vanishes: $V=0$. Moreover, the function~$s_0$ is defined by
\begin{equation}
  \label{eq:ParaStokes}
  s_0(r,\vartheta,\varphi,t) = \kappa \, K(\mathbf{r},t) \,
  \omega^{-\frac{p-1}{2}} \, \mathcal{D}_{\rm v}^{\frac{p+3}{2}}
  \, \left( \frac{B}{\Gamma_{\rm v}} \, \sqrt{ 1 - ( \mathcal{D}_{\rm v} \, \mathbf{n} \cdot \mathbf{b} )^2 } 
  \right)^{\frac{p+1}{2}}
\end{equation}
where $\omega$ is the angular frequency of the emitted radiation, and $\kappa$ is a constant factor that depends only on the nature of the radiating particles~(charge $q$ and mass $m$) and the power law index $p$ of their distribution
\begin{equation}
  \kappa = \frac{\sqrt{3}}{2\,\upi} \, \frac{1}{4} \, \Gamma_{\rm Eu} 
  \left( \frac{3\,p+7}{12}\right) \,
  \Gamma_{\rm Eu}\left(\frac{3\,p-1}{12}\right) \, 
  \frac{|q|^3}{4\,\upi\,\varepsilon_0\,m\,c} 
  \, \left( \frac{3\,|q|}{m^3\,c^4} \right)^{\frac{p-1}{2}} 
\end{equation}
with $\Gamma_{\rm Eu}$ the Euler gamma function. In the present study, the frequency $\omega$ is fixed and located between the lower and upper synchrotron cut-off frequency. The angle~$\psi$ measures the inclination of the local electric field with respect to the projection of the pulsar's rotation axis on the plane of the sky as seen in the observer's frame. The degree of linear polarization is defined by
\begin{equation}
 \Pi = \frac{\sqrt{Q^2 + U^2}}{I} . 
\end{equation}
The corresponding polarization angle, defined as the position angle between the electric field vector at the observer and the projection of the pulsar's rotation axis on the plane of the sky is then
\begin{equation}
 \psi = \frac{1}{2} \, \arctan \left( \frac{U}{Q} \right)  .
\end{equation}

\section{RESULTS}
\label{sec:Results}

We performed many calculations by changing each of the parameter of the model in a reasonable range of values. We compiled these results in an atlas containing hundreds of light-curves and associated polarization characteristics. It is obviously impossible to publish and show all these simulated data in a paper. Nevertheless, in order for the reader to get a flavour of the behaviour of these simulations, we will pick out some particular combinations of parameters which are relevant for understanding the evolution of polarization in the striped wind model. We now discuss the influence of these parameters on the polarization properties of the pulsed emission. The most important ones are
\begin{itemize}
 \item the region where emission is supposed to start~$r_0$.
 \item the constant Lorentz factor of the wind~$\Gamma_{\rm v}$.
 \item the ratio of cold to hot particle density number~$N_{\rm c}/N_{\rm h}$.
 \item the power-law index of the lepton distribution~$p$, let it be cold or hot.
 \item the pulsar obliquity~$\chi$.
 \item the inclination angle of the line of sight~$\zeta$.
\end{itemize}

\subsection{Location of the innermost emission region}

\begin{figure}
 \centering
 \includegraphics[width=0.5\textwidth]{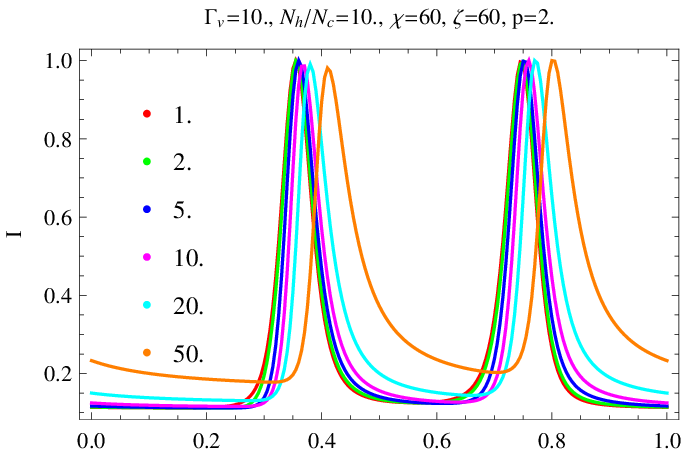} \\
 \includegraphics[width=0.5\textwidth]{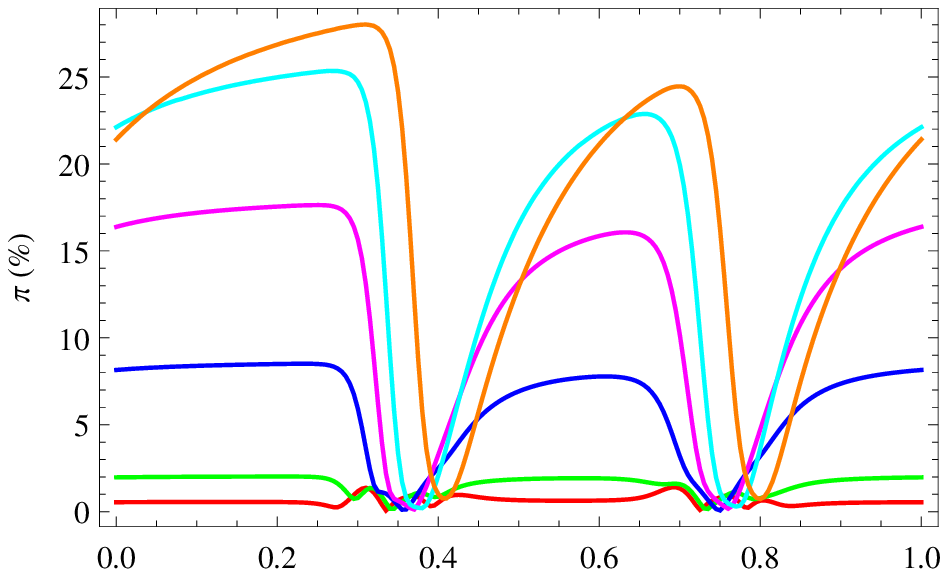} \\
 \includegraphics[width=0.5\textwidth]{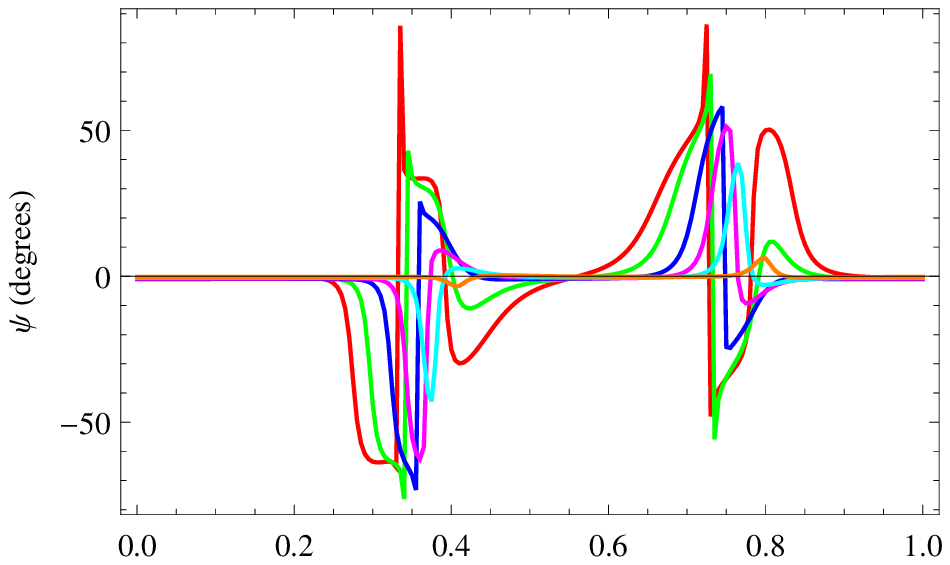} \\
 \caption{Evolution of the light-curves and polarization properties with respect to starting emission radius~$r_0$. The normalized intensity is shown in the upper panel, the polarization degree in the middle panel and the polarization angle in the lower panel.}
 \label{fig:emission_r0_1}
\end{figure}

We first look at the polarization properties depending on the location where emission starts, varying $r_0$ from $\rlight$ to $50\,\rlight$ choosing values in the set $\{1,2,5,10,20,50\}$. We indeed expect some variations of the emission characteristics because the electric drift speed and therefore the local frame where the electric field vanishes depends not only on $\vartheta$ but also on $r$. We show a typical example in fig.~\ref{fig:emission_r0_1}. In the following discussion, the specific set of parameters used to produce the plots are given in the frame title of the upper panel.  We see that the position where photons are produced only weakly influences the shape of the light-curves, upper panel of fig.~\ref{fig:emission_r0_1}. For the largest distances~$r_0$, the phase of the maximal intensity is slightly delayed compared to the smallest distances. There is a monotonic trend to increase the phase shift with $r_0$. This is explained by the delay introduced by the time of flight effect comparing two photons emitted at different radius but along the same line of sight. Indeed the difference in arrival time of two photons produced at a radius $r_1$ and $r_2$ respectively and normalized to the period~$P$ of the pulsar is given by
\begin{equation}
 \frac{\Delta t_{\rm a}}{P} = \left( \frac{1}{\beta_{\rm v}} - 1 \right) \, \frac{|r_2-r_1|}{2\,\upi\,\rlight} \approx \frac{|r_2-r_1|}{4\,\upi\,\Gamma_{\rm v}^2\,\rlight}.
\end{equation}
see for instance equation~(31) in \cite{2011MNRAS.412.1870P} and the discussion therein for more explanation. The approximation is valid for $\Gamma_{\rm v}\gg1$. In the figure, for $\Gamma_{\rm v}=10$, $r_1=\rlight$ and $r_2=50\,\rlight$ we get $\Delta t_{\rm a}/P\approx0.039$ in agreement with the light-curves. We checked that for $\Gamma_{\rm v}=50$ the same light-curves as those shown in figure~\ref{fig:emission_r0_1} almost overlap because the time lag is a factor 25 smaller thus not seen by eye. Note that both pulses are not symmetric with respect to rising and falling time. Next the polarization degree in the middle panel of fig.~\ref{fig:emission_r0_1} demonstrates an increase in polarization for large~$r_0$. Very close to the light-cylinder, there is no significant polarization whereas for $r_0>10\,\rlight$, the polarization degree can reach up to 28\% in the off-pulse phase in the particular case of $r_0=50\,\rlight$. In this phase, only the cold part of the wind contributes to the emissivity. Moreover this happens in a well-ordered magnetic field, therefore a high-polarization in accordance with a constant polarization angle as shown in the lower panel of fig.~\ref{fig:emission_r0_1}. The situation is drastically different in the pulses. Indeed, the contribution of the hot component depolarizes the synchrotron emission which is almost zero in the middle of the pulses. Meanwhile, the polarization angle suffers sharp gradient switching from -60/-80\degr to +60/+80\degr within each pulse. There exists no obvious symmetry in the polarization angle variation in both pulses. The phase range where the variation is significant is largest for small starting radii and diminishes for the largest~$r_0$.

\subsection{Lorentz factor}

\begin{figure}
 \centering
 \includegraphics[width=0.5\textwidth]{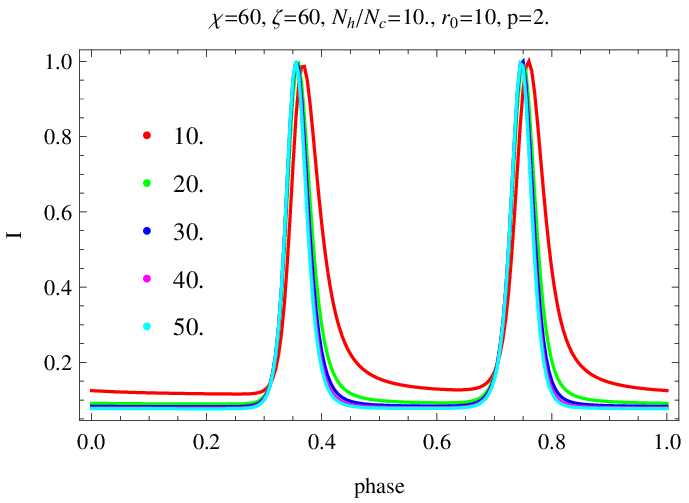} \\
 \includegraphics[width=0.5\textwidth]{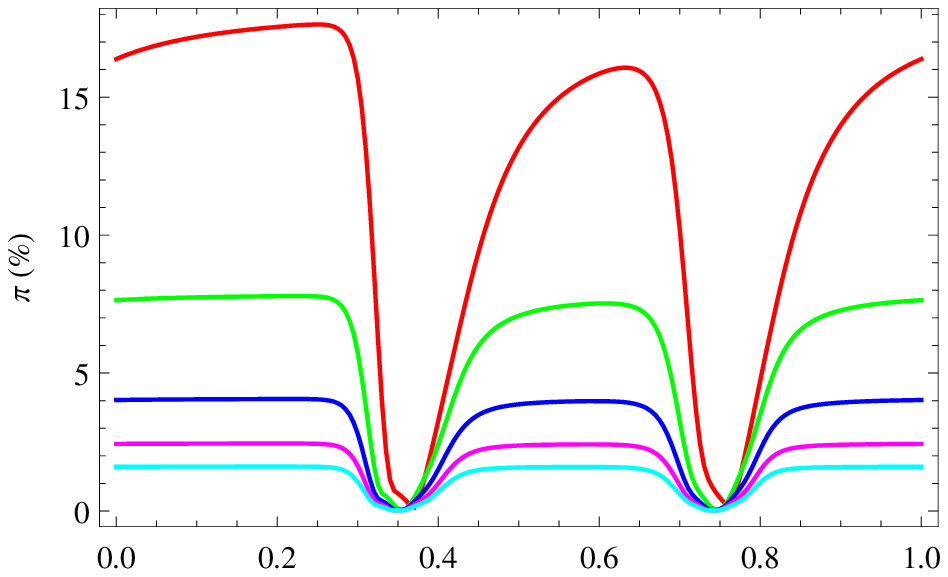} \\
 \includegraphics[width=0.5\textwidth]{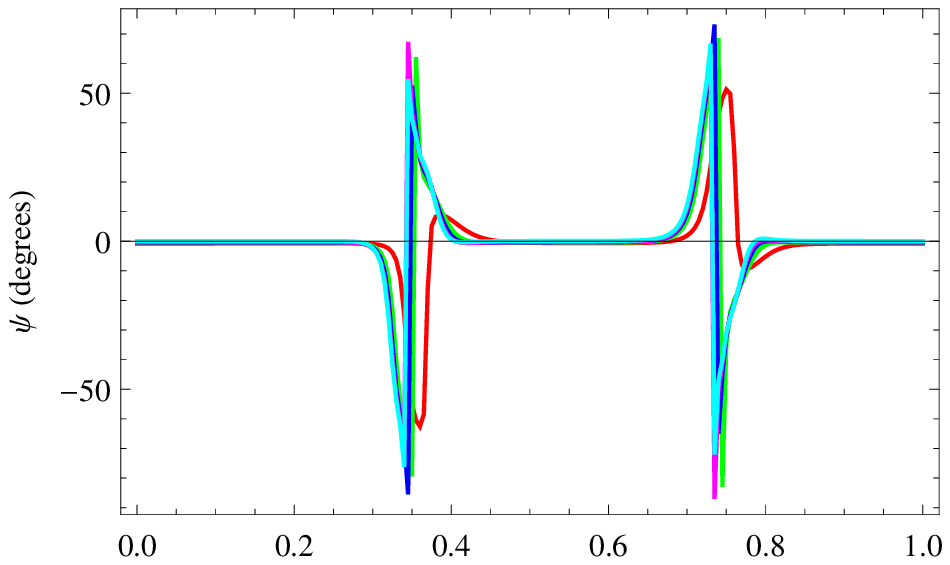} \\
 \caption{Evolution of the light-curves and polarization properties with respect to the Lorentz factor~$\Gamma_{\rm v}$.}
 \label{fig:pulse_gamma}
\end{figure}

The speed of the wind at its base is another badly constrained parameter. We show in fig.~\ref{fig:pulse_gamma} the influence of the Lorentz factor in the set $\{10,20,30,40,50\}$ on the polarization of the emission. For modest values of~$\Gamma_{\rm v}$, the light-curves are highly asymmetric, the falling time being longer than the rising time. The shape of the pulses is a combination of the current sheet thickness and the relativistic beaming effect as long as the opening angle of the beaming is larger than the thickness of the stripe, both rescaled according to the pulsar period, namely 360\degr = $2\,\upi\,\rlight$ = one wavelength = one period. For increasing $\Gamma_{\rm v}$, the opening angle decreases like $1/\Gamma_{\rm v}$, explaining the sharpening of the pulse profile for $\Gamma_{\rm v}=20$ compared to $\Gamma_{\rm v}=10$. At the point where the opening angle becomes comparable or even smaller than the thickness of the sheet, the pulse width decorrelates from the Lorentz factor and we see the current sheet thickness and nothing else. Thus, as expected, the width of the pulses tend to a minimum independent of the Lorentz factor for $\Gamma_{\rm v}\gg1$. This is indeed shown in the upper panel of fig.~\ref{fig:pulse_gamma}. Less intuitive is the corresponding polarization degree, middle panel of fig.~\ref{fig:pulse_gamma}. Low Lorentz factors correspond to high polarization degree, up to 18\% in the case shown. For higher Lorentz factors, $\Pi$ decreases quickly, attaining only 8\% for $\Gamma_{\rm v}=20$ and a bit less than 2\% for $\Gamma_{\rm v}=50$. It seems that relativistic flows depolarizes the synchrotron emission significantly in the striped wind. The polarization angle is only slightly affected by the variation of the Lorentz factor, lower panel of fig.~\ref{fig:pulse_gamma}.

\subsection{Particle density number}

\begin{figure}
 \centering
 \includegraphics[width=0.5\textwidth]{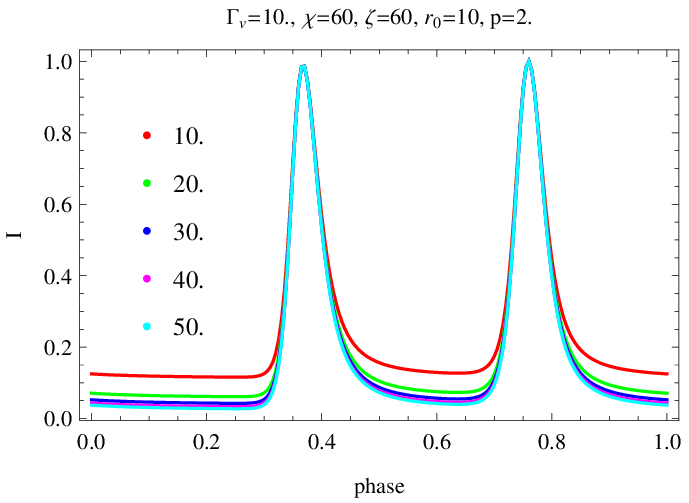} \\
 \includegraphics[width=0.5\textwidth]{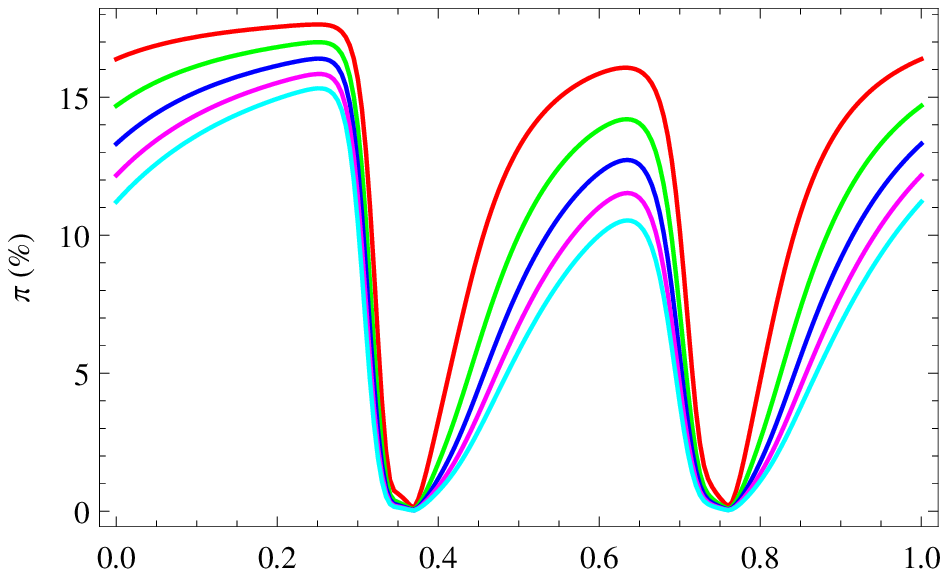} \\
 \includegraphics[width=0.5\textwidth]{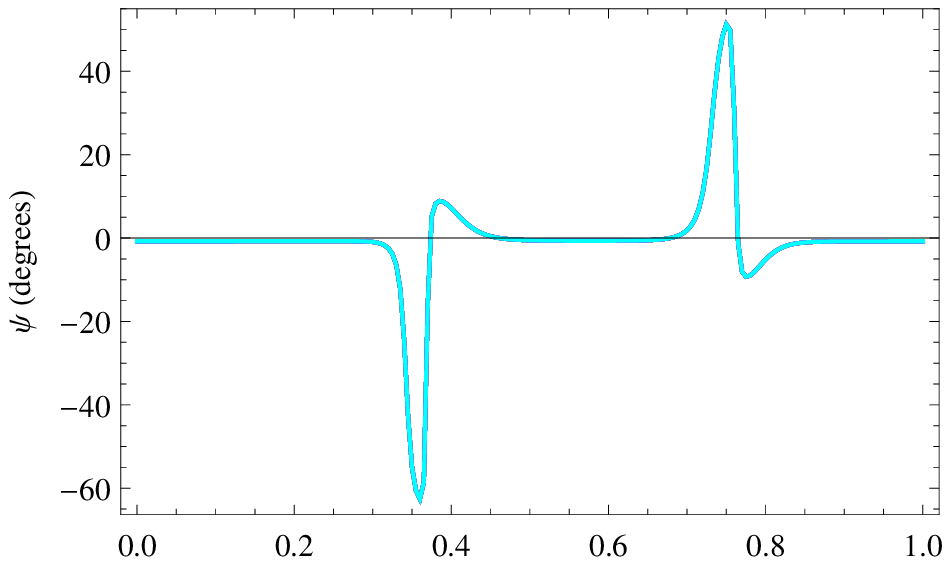} \\
 \caption{Evolution of the light-curves and polarization properties with respect to the particle density ratio $N_h/N_c$.}
 \label{fig:pulse_densite}
\end{figure}

The mass load in the wind is still uncertain. Additionally, the kinetic structure of the stripe, temperature and density profile as well as particle velocity remain unknown. Nevertheless, in order to estimate the behaviour of the polarization with respect to lepton populations, we investigate the change due to the particle density ratio $N_h/N_c$ chosen in the set $\{10,20,30,40,50\}$. An example is presented in fig.~\ref{fig:pulse_densite}. First we note that there is no distinction between the different calculations for the polarization angle, lower panel. The particle density number can not influence the angle. However, it influences the ratio between the peak intensity in the light-curve and the off-pulse emission. A higher density contrast, let us say $N_h/N_c=50$, induces a fainter off-pulse level compared to the case $N_h/N_c=10$, upper panel. This puts into picture the fact that the hot component is mainly associated with the pulses whereas the off-pulse part is essentially connected to the cold plasma component. The light-curves, whatever the density, do all overlap, all peaks are phase-aligned. The situation is identical for the polarization degree, its phase dependence looks similar for each curve. The main discrepancy lies in the highest polarization degree which decreases for increasing density contrast. The hot and unpolarized component reduces $\Pi$.

\subsection{Power-law index}

\begin{figure}
 \centering
 \includegraphics[width=0.5\textwidth]{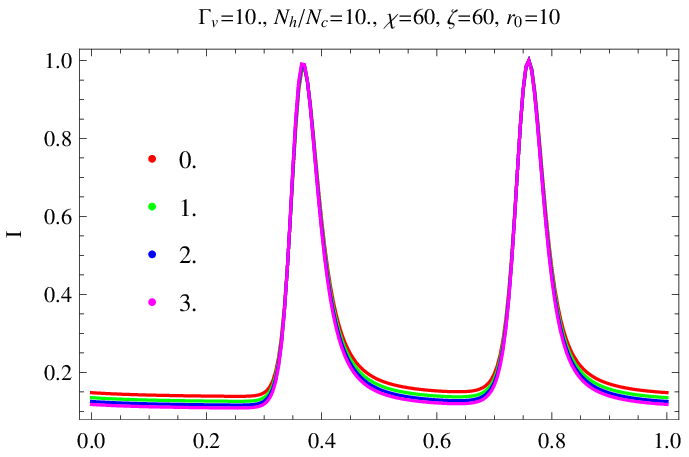} \\
 \includegraphics[width=0.5\textwidth]{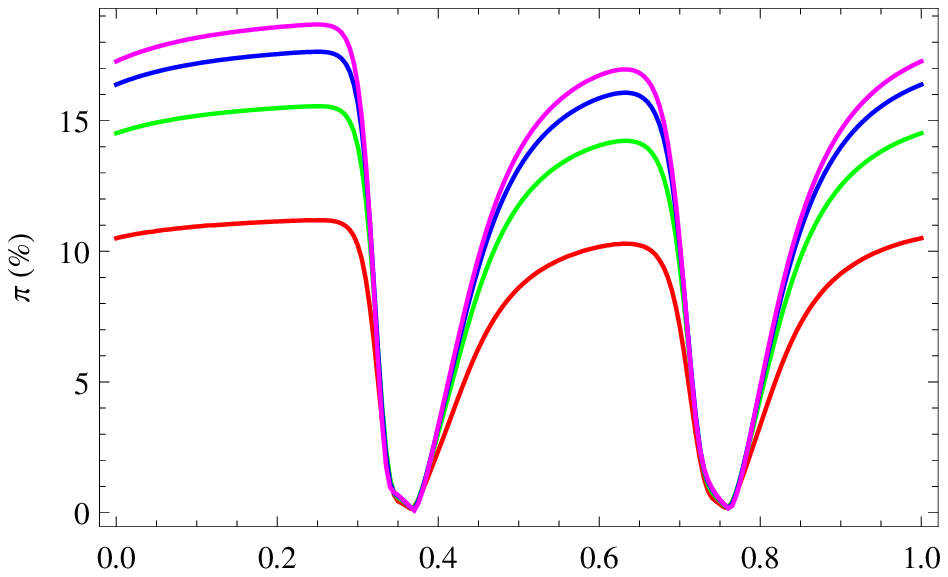} \\
 \includegraphics[width=0.5\textwidth]{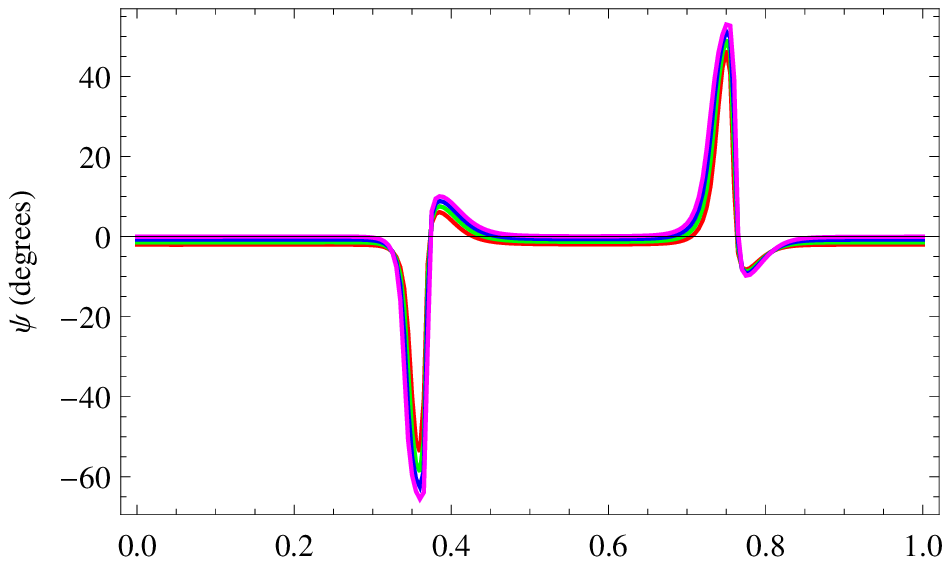} \\
 \caption{Evolution of the light-curves and polarization properties with respect to the power-law index~$p$.}
 \label{fig:emission_p_1}
\end{figure}

Related to the particle density number is the power-law index, actually both a part of the distribution function which is largely unknown. Thus we studied also the variations induced by $p$ chosen in the set $\{0,1,2,3\}$. The power-law index intervenes through the power in the Doppler beaming factor $\mathcal{D}_{\rm v}$ and renders the relativistic beaming more or less efficient depending on its precise value. This is seen in fig.~\ref{fig:emission_p_1}. The case $p=3$ reinforce the variation with $\mathcal{D}_{\rm v}$, efficient beaming implying less emission in the off-pulse phase compare to the others cases like $p=0$. The polarization degree follows a similar trend. Increasing $p$ also increases $\Pi$ substantially from 11\% to 19\%. Theoretically we know that the maximum possible degree of polarization is related to~$p$. Indeed, in the very simplistic and ideal case of a uniform and constant magnetic field, $\Pi = (p+1)/(p+7/3)$. For $p=0$ we have $\Pi=3/7\approx$42\% and for $p=3$ we get $\Pi=3/4$=75\% which is in the same ratio as 19\% and 11\% ($75/42\approx19/11$). The polarization angle, although differing from case to case, possesses a phase-resolved variation which is very similar for every value of $p$.

\subsection{Obliquity}

\begin{figure}
 \centering
 \includegraphics[width=0.5\textwidth]{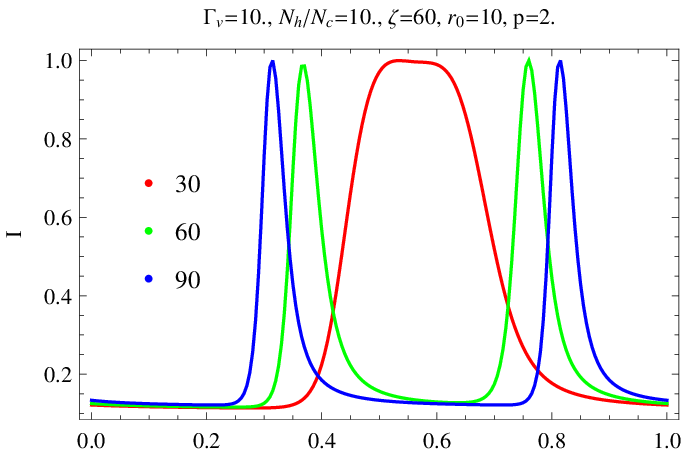} \\
 \includegraphics[width=0.5\textwidth]{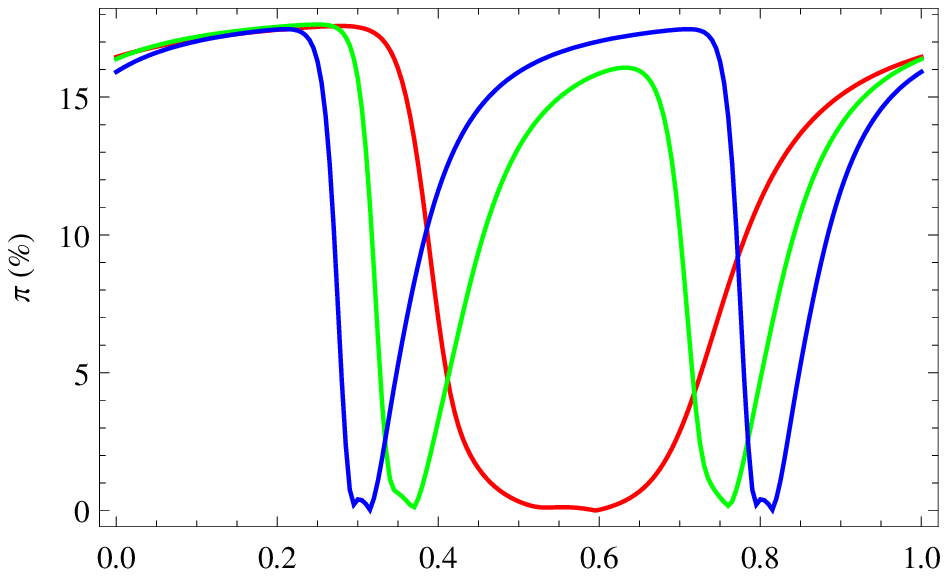} \\
 \includegraphics[width=0.5\textwidth]{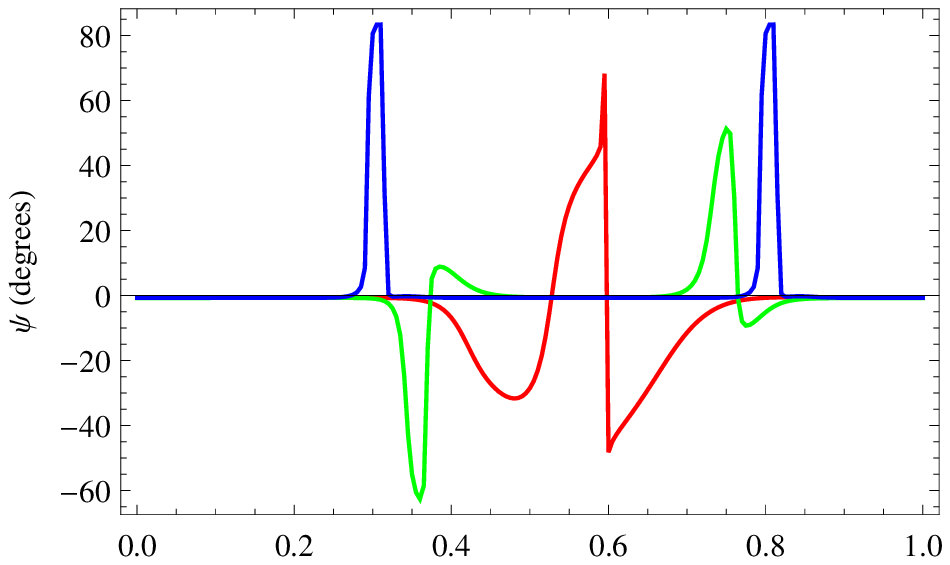} \\
 \caption{Evolution of the light-curves and polarization properties with respect to obliquity~$\chi$.}
 \label{fig:emission_chi_1}
\end{figure}

The two last parameters we want to study are geometrical, the obliquity and the inclination of the line of sight. For some pulsars, these orientations are well constrained like for the Crab for instance. Fig.~\ref{fig:emission_chi_1} shows an example of the polarization characteristics for different obliquities $\chi=\{30\degr,60\degr,90\degr\}$. For a small obliquity, in the case $\chi<\upi/2-\zeta$ the two peaks are not well separated and we observe a single pulse. This is the case for $\chi=30\degr$ and $\zeta=60\degr$ as shown in the upper panel of fig.~\ref{fig:emission_chi_1} because the line of sight does not significantly cross the stripe. On the contrary for $\chi=\{60\degr,90\degr\}$, the observer looks through the stripe and clearly distinguishes two pulses per period, their separation being maximal and equal to half a period for the orthogonal rotator $\chi=90\degr$. The polarization degree and angle evolve according to the shift between both pulses, if present. In any case, $\Pi$ decreases sharply within the pulse(s) and the angle shows sharp gradient. In the off-pulse phase, $\Pi$ is almost constant and maximal and the polarization angle is null, meaning that the electric field vector of the light is aligned with projection of the pulsar rotation axis onto the plane of the sky. The amplitude of the polarization angle sweep does not significantly depend on $\zeta$, although its location should be in phase with the pulses, lower panel.

\subsection{Line of sight inclination}

\begin{figure}
 \centering
 \includegraphics[width=0.5\textwidth]{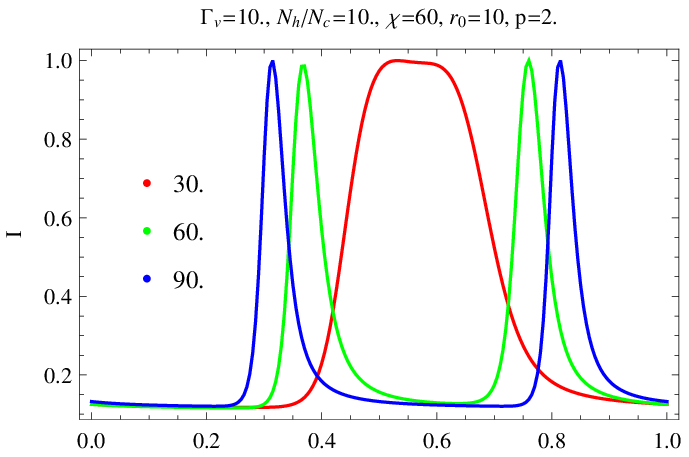} \\
 \includegraphics[width=0.5\textwidth]{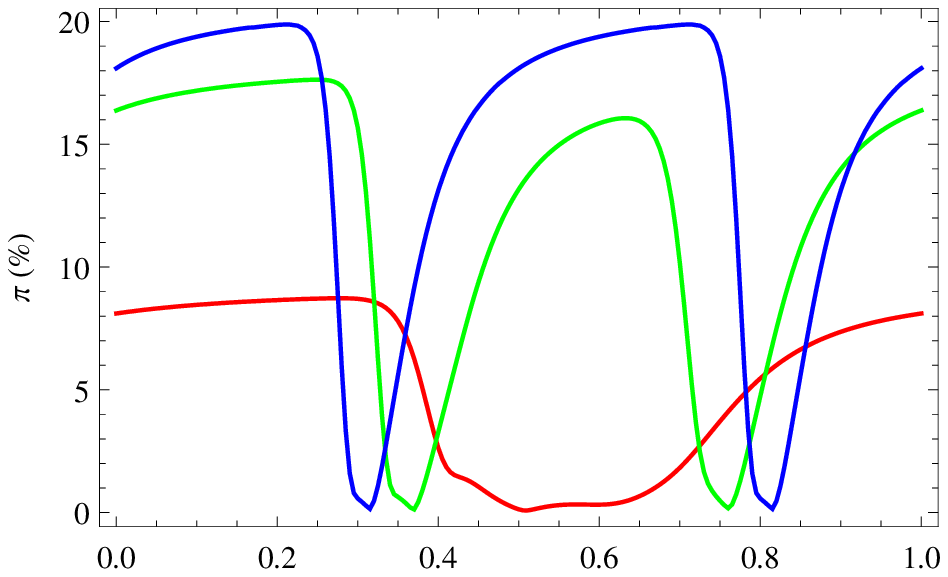} \\
 \includegraphics[width=0.5\textwidth]{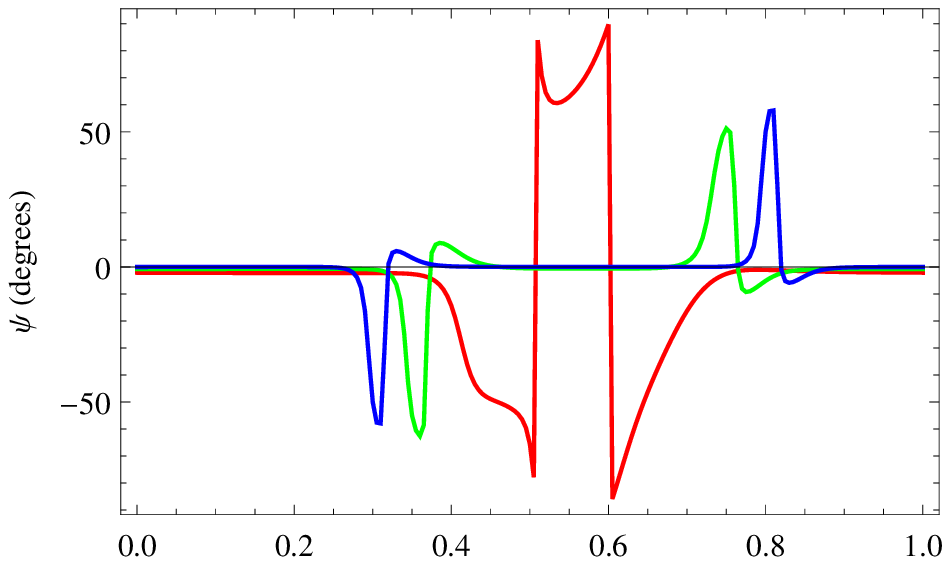} \\
 \caption{Evolution of the light-curves and polarization properties with respect to the inclination of the line of sight~$\zeta$.}
 \label{fig:emission_zeta_1}
\end{figure}

Finally, we looked for the influence solely of the inclination of the line of sight. Results are presented in fig.~\ref{fig:emission_zeta_1}. For the light-curves, shaped with one or two pulses, the same discussion as in the previous paragraph holds, two pulses are detected only if $\chi>\upi/2-\zeta$. From the middle panel, it is easily recognized that the polarization degree depends strongly on the inclination angle. Too a small $\zeta$ produces weakly polarized emission, for instance with $\zeta=30\degr$ wet get a maximum of 9\%. On the contrary, for maximal inclination with $\zeta=90\degr$ the polarized component is high, up to 20\%. Remember that close to the pole the striped wind is circularly polarized whereas close to the equatorial plane it is almost exclusively linearly polarized. The transition from a circularly to a linearly polarized wave when drifting from the rotation axis to the equator explains the evolution in polarization degree~$\Pi$. The amplitude of the polarization angle decreases significantly with increasing $\zeta$, lower panel.

\section{CONCLUSION}
\label{sec:Conclusion}

In the striped wind model, the high energy emission of pulsars, from infra-red to gamma-ray, arises from outside the light cylinder, in accordance with the suggestions of \cite{1970Natur.226..622P} and \cite{1970ApJ...159L..77S}. We tried to get constraints on the  arbitrary parameters concerning the configuration of the emission region and the distribution function of the emitting particles. Future observations of polarization properties in the high-energy band will be able to disentangle between several models, solving for the decade long standing problem about the emission mechanism and location. We presented detailed computations of the polarization properties of the pulses arising from the striped wind with finite thickness of the stripe. The electric vector of the off-pulse emission is aligned with the projection of the pulsar's rotation axis on the plane of the sky as was already demonstrated in \cite{2005ApJ...627L..37P}. This is in striking agreement with observations of the Crab pulsar and some sparse results for Vela \citep{2007A&A...467.1157M}. However a more quantitative analysis of the polarization in the pulsed  phase requires deep knowledge about the particle distribution functions. This task is tremendous as it encompasses small time and length scales into a macroscopic relativistic flow many orders of magnitude larger than the stripe. We are far from a self-consistent treatment of the global MHD flows including possibly magnetic reconnection and turbulence within the sheet. Current state of the art numerical computations are still unable to join the plasma kinetic regime to the fluid regime from the neutron star surface up to several tens or hundreds of light cylinder radii. The opportunity to get knowledge on the plasma in the wind on a theoretical basis seems far from reachable.


\begin{thebibliography}{42}
\expandafter\ifx\csname natexlab\endcsname\relax\def\natexlab#1{#1}\fi

\bibitem[{{Abdo} {et~al}\mbox{.}(2010){Abdo}, {Ackermann}, {Ajello}, {Atwood},
  {Axelsson}, {Baldini}, {Ballet}, {Barbiellini}, {Baring}, {Bastieri},
  {Baughman}, {Bechtol}, {Bellazzini}, {Berenji}, {Blandford}, {Bloom},
  {Bonamente}, {Borgland}, {Bregeon}, {Brez}, {Brigida}, {Bruel}, {Burnett},
  {Buson}, {Caliandro}, {Cameron}, {Camilo}, {Caraveo}, {Casandjian}, {Cecchi},
  {{\c C}elik}, {Charles}, {Chekhtman}, {Cheung}, {Chiang}, {Ciprini}, {Claus},
  {Cognard}, {Cohen-Tanugi}, {Cominsky}, {Conrad}, {Corbet}, {Cutini}, {den
  Hartog}, {Dermer}, {de Angelis}, {de Luca}, {de Palma}, {Digel}, {Dormody},
  {Silva}, {Drell}, {Dubois}, {Dumora}, {Espinoza}, {Farnier}, {Favuzzi},
  {Fegan}, {Ferrara}, {Focke}, {Fortin}, {Frailis}, {Freire}, {Fukazawa},
  {Funk}, {Fusco}, {Gargano}, {Gasparrini}, {Gehrels}, {Germani}, {Giavitto},
  {Giebels}, {Giglietto}, {Giommi}, {Giordano}, {Glanzman}, {Godfrey},
  {Gotthelf}, {Grenier}, {Grondin}, {Grove}, {Guillemot}, {Guiriec}, {Gwon},
  {Hanabata}, {Harding}, {Hayashida}, {Hays}, {Hughes}, {Jackson},
  {J{\'o}hannesson}, {Johnson}, {Johnson}, {Johnson}, {Johnson}, {Johnston},
  {Kamae}, {Kanbach}, {Kaspi}, {Katagiri}, {Kataoka}, {Kawai}, {Kerr},
  {Kn{\"o}dlseder}, {Kocian}, {Kramer}, {Kuss}, {Lande}, {Latronico},
  {Lemoine-Goumard}, {Livingstone}, {Longo}, {Loparco}, {Lott}, {Lovellette},
  {Lubrano}, {Lyne}, {Madejski}, {Makeev}, {Manchester}, {Marelli},
  {Mazziotta}, {McConville}, {McEnery}, {McGlynn}, {Meurer}, {Michelson},
  {Mineo}, {Mitthumsiri}, {Mizuno}, {Moiseev}, {Monte}, {Monzani}, {Morselli},
  {Moskalenko}, {Murgia}, {Nakamori}, {Nolan}, {Norris}, {Noutsos}, {Nuss},
  {Ohsugi}, {Omodei}, {Orlando}, {Ormes}, {Ozaki}, {Paneque}, {Panetta},
  {Parent}, {Pelassa}, {Pepe}, {Pesce-Rollins}, {Piron}, {Porter}, {Rain{\`o}},
  {Rando}, {Ransom}, {Ray}, {Razzano}, {Rea}, {Reimer}, {Reimer}, {Reposeur},
  {Ritz}, {Rodriguez}, {Romani}, {Roth}, {Ryde}, {Sadrozinski}, {Sanchez},
  {Sander}, {Saz Parkinson}, {Scargle}, {Schalk}, {Sellerholm}, {Sgr{\`o}},
  {Siskind}, {Smith}, {Smith}, {Spandre}, {Spinelli}, {Stappers}, {Starck},
  {Striani}, {Strickman}, {Strong}, {Suson}, {Tajima}, {Takahashi},
  {Takahashi}, {Tanaka}, {Thayer}, {Thayer}, {Theureau}, {Thompson},
  {Thorsett}, {Tibaldo}, {Tibolla}, {Torres}, {Tosti}, {Tramacere}, {Uchiyama},
  {Usher}, {Van Etten}, {Vasileiou}, {Venter}, {Vilchez}, {Vitale}, {Waite},
  {Wang}, {Wang}, {Watters}, {Weltevrede}, {Winer}, {Wood}, {Ylinen}, \&
  {Ziegler}}]{2010ApJS..187..460A}
{Abdo} A.~A. {et~al.}, 2010, \apjs, 187, 460

\bibitem[{{Aliu} {et~al}\mbox{.}(2008){Aliu}, {Anderhub}, {Antonelli},
  {Antoranz}, {Backes}, {Baixeras}, {Barrio}, {Bartko}, {Bastieri}, {Becker},
  {Bednarek}, {Berger}, {Bernardini}, {Bigongiari}, {Biland}, {Bock},
  {Bonnoli}, {Bordas}, {Bosch-Ramon}, {Bretz}, {Britvitch}, {Camara},
  {Carmona}, {Chilingarian}, {Commichau}, {Contreras}, {Cortina}, {Costado},
  {Covino}, {Curtef}, {Dazzi}, {De Angelis}, {De Cea del Pozo}, {de los Reyes},
  {De Lotto}, {De Maria}, {De Sabata}, {Delgado Mendez}, {Dominguez}, {Dorner},
  {Doro}, {Els{\"a}sser}, {Errando}, {Fagiolini}, {Ferenc}, {Fernandez},
  {Firpo}, {Fonseca}, {Font}, {Galante}, {Garcia Lopez}, {Garczarczyk}, {Gaug},
  {Goebel}, {Hadasch}, {Hayashida}, {Herrero}, {H{\"o}hne}, {Hose}, {Hsu},
  {Huber}, {Jogler}, {Kranich}, {La Barbera}, {Laille}, {Leonardo}, {Lindfors},
  {Lombardi}, {Longo}, {Lopez}, {Lorenz}, {Majumdar}, {Maneva}, {Mankuzhiyil},
  {Mannheim}, {Maraschi}, {Mariotti}, {Martinez}, {Mazin}, {Meucci}, {Meyer},
  {Miranda}, {Mirzoyan}, {Moles}, {Moralejo}, {Nieto}, {Nilsson}, {Ninkovic},
  {Otte}, {Oya}, {Paoletti}, {Paredes}, {Pasanen}, {Pascoli}, {Pauss}, {Pegna},
  {Perez-Torres}, {Persic}, {Peruzzo}, {Piccioli}, {Prada}, {Prandini},
  {Puchades}, {Raymers}, {Rhode}, {Rib{\'o}}, {Rico}, {Rissi}, {Robert},
  {R{\"u}gamer}, {Saggion}, {Saito}, {Salvati}, {Sanchez-Conde}, {Sartori},
  {Satalecka}, {Scalzotto}, {Scapin}, {Schweizer}, {Shayduk}, {Shinozaki},
  {Shore}, {Sidro}, {Sierpowska-Bartosik}, {Sillanp{\"a}{\"a}}, {Sobczynska},
  {Spanier}, {Stamerra}, {Stark}, {Takalo}, {Tavecchio}, {Temnikov}, {Tescaro},
  {Teshima}, {Tluczykont}, {Torres}, {Turini}, {Vankov}, {Venturini}, {Vitale},
  {Wagner}, {Wittek}, {Zabalza}, {Zandanel}, {Zanin}, {Zapatero}, {de Jager},
  {de Ona Wilhelmi}, \& {MAGIC Collaboration}}]{2008Sci...322.1221A}
{Aliu} E. {et~al.}, 2008, Science, 322, 1221

\bibitem[{{Campana} {et~al}\mbox{.}(2009){Campana}, {Massaro}, {Mineo}, \&
  {Cusumano}}]{2009A&A...499..847C}
{Campana} R., {Massaro} E., {Mineo} T., {Cusumano} G., 2009, \aap, 499, 847

\bibitem[{{Cheng} {et~al}\mbox{.}(1986){Cheng}, {Ho}, \&
  {Ruderman}}]{1986ApJ...300..500C}
{Cheng} K.~S., {Ho} C., {Ruderman} M., 1986, \apj, 300, 500

\bibitem[{{Cocke} {et~al}\mbox{.}(1970){Cocke}, {Disney}, \&
  {Muncaster}}]{1970Natur.227.1327C}
{Cocke} W.~J., {Disney} M.~J., {Muncaster} G.~W., 1970, \nat, 227, 1327

\bibitem[{{Cocke} {et~al}\mbox{.}(1973){Cocke}, {Ferguson}, \&
  {Muncaster}}]{1973ApJ...183..987C}
{Cocke} W.~J., {Ferguson} D.~C., {Muncaster} G.~W., 1973, \apj, 183, 987

\bibitem[{{Coroniti}(1990)}]{1990ApJ...349..538C}
{Coroniti} F.~V., 1990, \apj, 349, 538

\bibitem[{{Crusius-W{\"a}tzel} {et~al}\mbox{.}(2001){Crusius-W{\"a}tzel},
  {Kunzl}, \& {Lesch}}]{2001ApJ...546..401C}
{Crusius-W{\"a}tzel} A.~R., {Kunzl} T., {Lesch} H., 2001, \apj, 546, 401

\bibitem[{{Dean} {et~al}\mbox{.}(2008){Dean}, {Clark}, {Stephen}, {McBride},
  {Bassani}, {Bazzano}, {Bird}, {Hill}, {Shaw}, \&
  {Ubertini}}]{2008Sci...321.1183D}
{Dean} A.~J. {et~al.}, 2008, Science, 321, 1183

\bibitem[{{Dyks} {et~al}\mbox{.}(2004){Dyks}, {Harding}, \&
  {Rudak}}]{2004ApJ...606.1125D}
{Dyks} J., {Harding} A.~K., {Rudak} B., 2004, \apj, 606, 1125

\bibitem[{{Dyks} \& {Rudak}(2003)}]{2003ApJ...598.1201D}
{Dyks} J., {Rudak} B., 2003, \apj, 598, 1201

\bibitem[{{Ferguson}(1973)}]{1973ApJ...183..977F}
{Ferguson} D.~C., 1973, \apj, 183, 977

\bibitem[{{Ferguson} {et~al}\mbox{.}(1974){Ferguson}, {Cocke}, \&
  {Gehrels}}]{1974ApJ...190..375F}
{Ferguson} D.~C., {Cocke} W.~J., {Gehrels} T., 1974, \apj, 190, 375

\bibitem[{{Forot} {et~al}\mbox{.}(2008){Forot}, {Laurent}, {Grenier},
  {Gouiff{\`e}s}, \& {Lebrun}}]{2008ApJ...688L..29F}
{Forot} M., {Laurent} P., {Grenier} I.~A., {Gouiff{\`e}s} C., {Lebrun} F.,
  2008, \apjl, 688, L29

\bibitem[{{Graham-Smith} {et~al}\mbox{.}(1996){Graham-Smith}, {Dolan}, {Boyd},
  {Biggs}, {Lyne}, \& {Percival}}]{1996MNRAS.282.1354G}
{Graham-Smith} F., {Dolan} J.~F., {Boyd} P.~T., {Biggs} J.~D., {Lyne} A.~G.,
  {Percival} J.~W., 1996, \mnras, 282, 1354

\bibitem[{{Jones} {et~al}\mbox{.}(1981){Jones}, {Smith}, \&
  {Wallace}}]{1981MNRAS.196..943J}
{Jones} D.~H.~P., {Smith} F.~G., {Wallace} P.~T., 1981, \mnras, 196, 943

\bibitem[{{Kirk} \& {Skj{\ae}raasen}(2003)}]{2003ApJ...591..366K}
{Kirk} J.~G., {Skj{\ae}raasen} O., 2003, \apj, 591, 366

\bibitem[{{Kirk} {et~al}\mbox{.}(2002){Kirk}, {Skj{\ae}raasen}, \&
  {Gallant}}]{2002A&A...388L..29K}
{Kirk} J.~G., {Skj{\ae}raasen} O., {Gallant} Y.~A., 2002, \aap, 388, L29

\bibitem[{{Kristian} {et~al}\mbox{.}(1970){Kristian}, {Visvanathan},
  {Westphal}, \& {Snellen}}]{1970ApJ...162..475K}
{Kristian} J., {Visvanathan} N., {Westphal} J.~A., {Snellen} G.~H., 1970, \apj,
  162, 475

\bibitem[{{Lyubarsky} \& {Kirk}(2001)}]{2001ApJ...547..437L}
{Lyubarsky} Y., {Kirk} J.~G., 2001, \apj, 547, 437

\bibitem[{{Lyutikov} {et~al}\mbox{.}(2003){Lyutikov}, {Pariev}, \&
  {Blandford}}]{2003ApJ...597..998L}
{Lyutikov} M., {Pariev} V.~I., {Blandford} R.~D., 2003, \apj, 597, 998

\bibitem[{{Massaro} {et~al}\mbox{.}(2006){Massaro}, {Campana}, {Cusumano}, \&
  {Mineo}}]{2006A&A...459..859M}
{Massaro} E., {Campana} R., {Cusumano} G., {Mineo} T., 2006, \aap, 459, 859

\bibitem[{{Michel}(1994)}]{1994ApJ...431..397M}
{Michel} F.~C., 1994, \apj, 431, 397

\bibitem[{{Mignani} {et~al}\mbox{.}(2007){Mignani}, {Bagnulo}, {Dyks}, {Lo
  Curto}, \& {S{\l}owikowska}}]{2007A&A...467.1157M}
{Mignani} R.~P., {Bagnulo} S., {Dyks} J., {Lo Curto} G., {S{\l}owikowska} A.,
  2007, \aap, 467, 1157

\bibitem[{{Nolan} {et~al}\mbox{.}(2012){Nolan}, {Abdo}, {Ackermann}, {Ajello},
  {Allafort}, {Antolini}, {Atwood}, {Axelsson}, {Baldini}, {Ballet}, \&
  et~al.}]{2012ApJS..199...31N}
{Nolan} P.~L. {et~al.}, 2012, \apjs, 199, 31

\bibitem[{{Pacini}(1970)}]{1970Natur.226..622P}
{Pacini} F., 1970, \nat, 226, 622

\bibitem[{{P{\'e}tri}(2009)}]{2009A&A...503...13P}
{P{\'e}tri} J., 2009, \aap, 503, 13

\bibitem[{{P{\'e}tri}(2011)}]{2011MNRAS.412.1870P}
{P{\'e}tri} J., 2011, \mnras, 412, 1870

\bibitem[{{P{\'e}tri}(2012)}]{2012MNRAS.424.2023P}
{P{\'e}tri} J., 2012, \mnras, 424, 2023

\bibitem[{{P{\'e}tri} \& {Kirk}(2005)}]{2005ApJ...627L..37P}
{P{\'e}tri} J., {Kirk} J.~G., 2005, \apjl, 627, L37

\bibitem[{{Romani} \& {Watters}(2010)}]{2010ApJ...714..810R}
{Romani} R.~W., {Watters} K.~P., 2010, \apj, 714, 810

\bibitem[{{Romani} \& {Yadigaroglu}(1995)}]{1995ApJ...438..314R}
{Romani} R.~W., {Yadigaroglu} I.-A., 1995, \apj, 438, 314

\bibitem[{{Ruderman} \& {Sutherland}(1975)}]{1975ApJ...196...51R}
{Ruderman} M.~A., {Sutherland} P.~G., 1975, \apj, 196, 51

\bibitem[{{Shklovsky}(1970)}]{1970ApJ...159L..77S}
{Shklovsky} I.~S., 1970, \apjl, 159, L77

\bibitem[{{Silver} {et~al}\mbox{.}(1978){Silver}, {Kestenbaum}, {Long},
  {Novick}, {Wolff}, \& {Weisskopf}}]{1978ApJ...225..221S}
{Silver} E.~H., {Kestenbaum} H.~L., {Long} K.~S., {Novick} R., {Wolff} R.~S.,
  {Weisskopf} M.~C., 1978, \apj, 225, 221

\bibitem[{{S{\l}owikowska} {et~al}\mbox{.}(2009){S{\l}owikowska}, {Kanbach},
  {Kramer}, \& {Stefanescu}}]{2009MNRAS.397..103S}
{S{\l}owikowska} A., {Kanbach} G., {Kramer} M., {Stefanescu} A., 2009, \mnras,
  397, 103

\bibitem[{{Smith} {et~al}\mbox{.}(1988){Smith}, {Jones}, {Dick}, \&
  {Pike}}]{1988MNRAS.233..305S}
{Smith} F.~G., {Jones} D.~H.~P., {Dick} J.~S.~B., {Pike} C.~D., 1988, \mnras,
  233, 305

\bibitem[{{Sturrock}(1970)}]{1970Natur.227..465S}
{Sturrock} P.~A., 1970, \nat, 227, 465

\bibitem[{{Takata} \& {Chang}(2007)}]{2007ApJ...670..677T}
{Takata} J., {Chang} H.-K., 2007, \apj, 670, 677

\bibitem[{{Tang} {et~al}\mbox{.}(2008){Tang}, {Takata}, {Jia}, \&
  {Cheng}}]{2008ApJ...676..562T}
{Tang} A.~P.~S., {Takata} J., {Jia} J.~J., {Cheng} K.~S., 2008, \apj, 676, 562

\bibitem[{{Venter} {et~al}\mbox{.}(2009){Venter}, {Harding}, \&
  {Guillemot}}]{2009ApJ...707..800V}
{Venter} C., {Harding} A.~K., {Guillemot} L., 2009, \apj, 707, 800

\bibitem[{{Wampler} {et~al}\mbox{.}(1969){Wampler}, {Scargle}, \&
  {Miller}}]{1969ApJ...157L...1W}
{Wampler} E.~J., {Scargle} J.~D., {Miller} J.~S., 1969, \apjl, 157, L1

\end{thebibliography}

\end{document}